\documentclass[12pt]{article}
\usepackage{graphicx}
\usepackage{lineno}
\usepackage{amsmath}
\usepackage{siunitx}
\usepackage{booktabs}
\usepackage[sorting=none]{biblatex}
\newcommand\pubnumber{}
\newcommand\pubdate{\today}

\def\institute{Physikalisches Institut\\
Rheinische Friedrich-Wilhelms-Universit\"{a}t Bonn, GERMANY}
\def\support{\footnote{Speaker}}

\def\Title#1{\begin{center} {\Large #1 } \end{center}}
\def\Author#1{\begin{center}{ \sc #1} \end{center}}
\def\Address#1{\begin{center}{ \it #1} \end{center}}

\newcommand\pubblock{\rightline{\begin{tabular}{l} \pubnumber\\
         \pubdate  \end{tabular}}}
\newenvironment{Abstract}{\begin{quotation}  }{\end{quotation}}
\newenvironment{Presented}{\begin{quotation} \begin{center} 
             PRESENTED AT\end{center}\bigskip 
      \begin{center}\begin{large}}{\end{large}\end{center} \end{quotation}}

\begin{document}
\begin{titlepage}
	
\pubblock

%\linenumbers
\vfill
\Title{Measurement of the production cross-section of a single top-quark in association with a $Z$ boson at 13 TeV with the ATLAS detector}
\vfill
\Author{ Irina Cioar\u{a}\support , on behalf of the ATLAS Collaboration}
\Address{\institute}
\vfill
\begin{Abstract}

The production of a top quark in association with a $Z$ boson is studied in the trilepton channel.
The data collected by the ATLAS experiment at the LHC in 2015 and 2016
at a centre-of-mass energy of $\sqrt{s} = 13\,\text{TeV}$ are used,
corresponding to an integrated luminosity of $36.1\,\text{fb}^{-1}$.
Events containing three identified leptons (electron and/or muon) and
two jets, one of which is identified as a $b$-quark jet are selected.
The major backgrounds come from diboson, $t\bar{t}$ and $Z\text{\,+\,jets}$ production.
A neural network is used to improve the background rejection and extract the signal. 
The resulting significance of the signal is $4.2\sigma$ in the data and the expected significance is $5.4\sigma$.
The measured cross-section is $600 \pm 170\text{(stat)} \pm 140\text{(syst)}\,\text{fb}$.

\end{Abstract}
\vfill
\begin{Presented}
$10^{th}$ International Workshop on Top Quark Physics\\
Braga, Portugal,  September 17--22, 2017
\end{Presented}
\vfill
\end{titlepage}
\def\thefootnote{\fnsymbol{footnote}}
\setcounter{footnote}{0}
%
%\linenumbers
\section{Introduction}

\begin{figure}[htbp]
	\centering
	\begin{tabular}{cc}
		\raisebox{0ex}{\includegraphics[width=.3\textwidth]{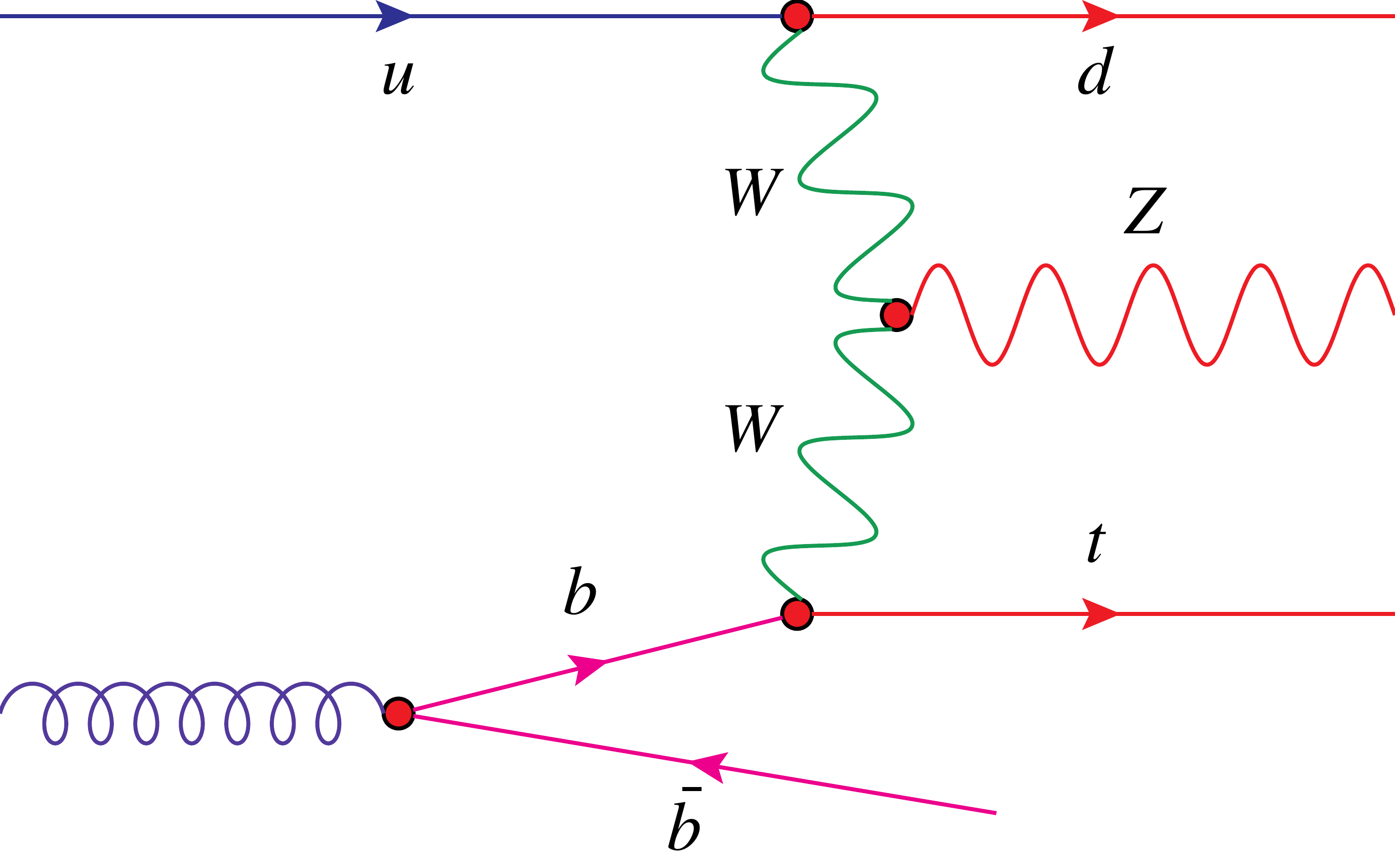}} & 
		\raisebox{1.2ex}{\includegraphics[width=.3\textwidth]{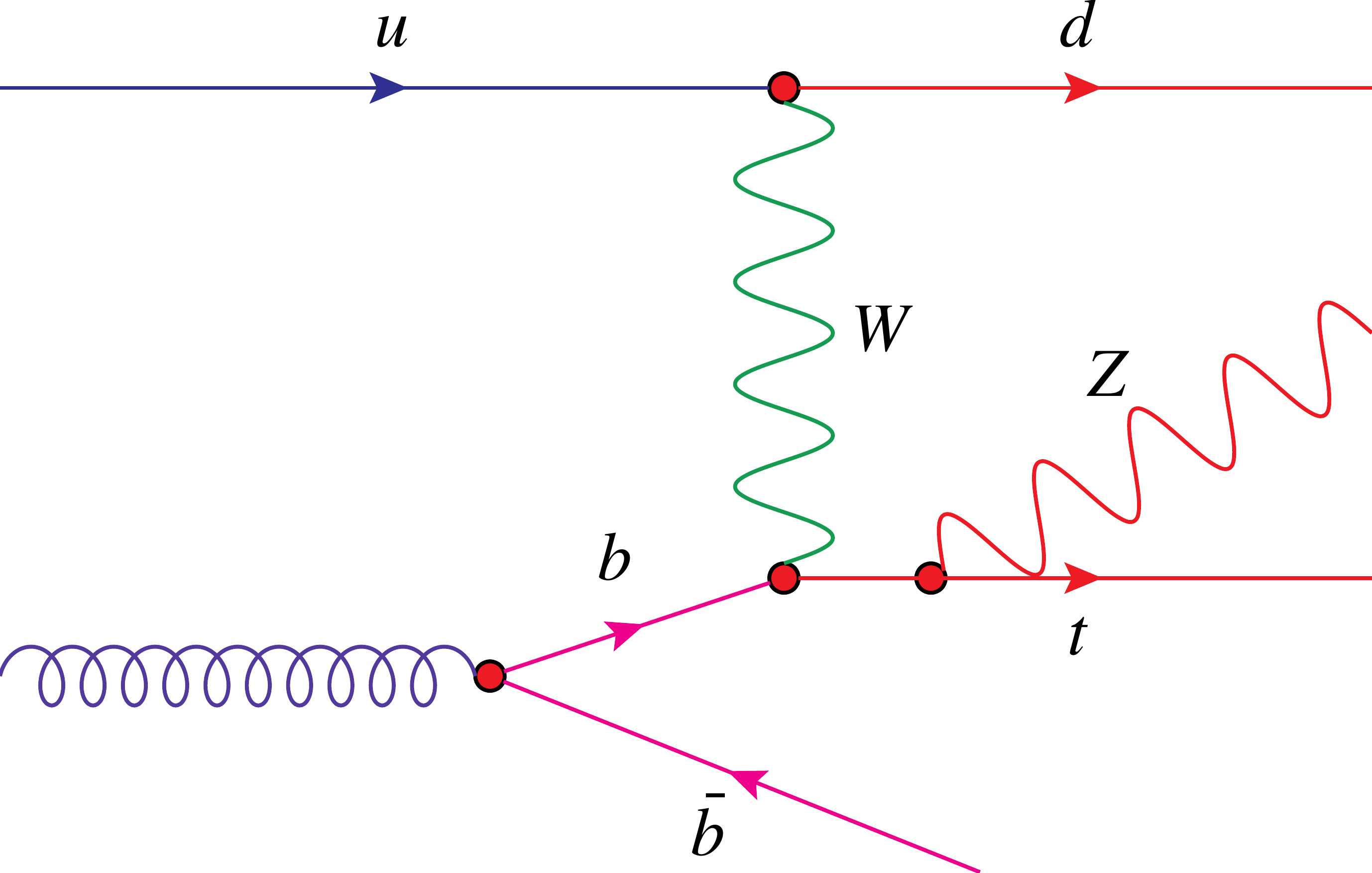}} \\
		%\raisebox{5ex}{\includegraphics[width=.23\textwidth]{tZq_4FS_feyn6}}\\
	\end{tabular}
	\caption{Example Feynman diagrams of the lowest-order
		amplitudes for the $tZq$ process. In the four-flavour scheme, the $b$-quark originates from gluon splitting \cite{Aaboud:2017ylb}.}% The wavy line denotes a \PW or \PZ/$\gamma^{*}$ boson~\protect\cite{Campbell:2013yla}.}
	\label{fig:feynman}
\end{figure}

Since its discovery in 1995, the top quark has been studied extensively using hadron collision data. With the latest increase in energy and instantaneous luminosity, the Large Hadron Collider (LHC) has made it possible to study top-quark processes that have a very low production cross-section. One of them is the associated production of a top quark and a $Z$ boson ($tZq$ production) \cite{Campbell:2013yla}. This occurs through the radiation of a $Z$ boson from any of the quark lines of $t$-channel, single top-quark production or through a three-boson coupling. Example Feynman diagrams of the lowest-order amplitudes are shown in Figure \ref{fig:feynman}. Hence, this channel is sensitive to both the coupling of the top to the $Z$, as well as the $WWZ$ coupling. This document presents a search for such rare events \cite{Aaboud:2017ylb}, performed using 2015-2016 collision data collected by the ATLAS detector \cite{PERF-2007-01} at a centre-of-mass energy of \SI{13}{\TeV}. The predicted production cross-section for $tZq$ events at next-to-leading order (NLO) in QCD is $\sigma_{tZq}^{\text{pred.}}=800^{+6.1\%}_{-7.4\%}\text{ fb}$.

\section{Analysis strategy}
The final state that offers the best potential for discovery (and highest signal to background ratio) is the one in which both the $Z$ boson and the $W$ boson coming from the top quark decay leptonically. This results in three leptons, two jets, one of which has to be identified as coming from a $b$ quark ($b$-tagged), and missing transverse momentum. Based on that, cuts are applied in order to define a signal region (SR) enriched in $tZq$ events. 
The number of signal events is estimated using a Monte Carlo (MC) simulated sample that is generated at leading order (LO) in QCD that is then normalised to the NLO predicted cross-section. 

The backgrounds coming from other sources that can result in similar final states are estimated either from MC samples or using data-driven methods. The modelling of these backgrounds is checked in dedicated validation regions (VR). 

Multivariate analysis techniques are used to construct a single discriminant that separates signal from background. This distribution is then fitted using a binned likelihood fit in order to extract the measured $tZq$ cross-section.  

\subsection{Event Selection and background estimation}

The cuts applied in order to select events that match the $tZq$ signature target the trileptonic final state described above and are summarised in Table \ref{tab:cuts}. Out of the three isolated leptons required, two of them must form an opposite sign, same flavour (OSSF) pair with a mass within a \SI{10}{\GeV} window around the $Z$ boson mass. One of the two selected jets must pass $b$-tagging criteria and be reconstructed in the central region of the detector ($|\eta|<2.5$), while the remaining light jet, that is expected to go mostly in the forward direction, can have $|\eta|$ up to \SI{4.5}{}. 

\begin{table}[htbp]
	\centering
	\begin{tabular}{ccc}
		\toprule
		 \multicolumn{3}{c}{Common selections}  \\
		\midrule
		\multicolumn{3}{c}{Exactly 3 leptons with $|\eta| < 2.5$ and $p_T> \SI{15}{\GeV}$}    \\
		\multicolumn{3}{c}{$p_T(\ell_1)> \SI{28}{\GeV}$, $p_T(\ell_2)> \SI{25}{\GeV}$, $p_T(\ell_3)> \SI{15}{\GeV}$}   \\
	 \multicolumn{3}{c}{$p_T(\text{jet})> \SI{30}{\GeV}$}   \\
		 \multicolumn{3}{c}{$m_T(\ell_{W},\nu) > \SI{20}{\GeV}$} \\
		\midrule
		SR & Diboson VR / CR & $t\bar{t}$ VR  \\
		\midrule
		$\ge$ 1 OSSF pair & $\ge$ 1 OSOF pair & $\ge$ 1 OSOF pair \\
		$|m_{ll} - m_Z| <  \SI{10}{\GeV}$ & $|m_{ll} - m_Z| < $ \SI{10}{\GeV} &$|m_{ll} - m_Z| > \SI{10}{\GeV}$ \\
		%    = 2 jets with  $|\eta| < $ 4.5  & = 1 jet with  $|\eta| < $ 4.5  & = 2 jets with  $|\eta| < $ 4.5  & = 2 jets with  $|\eta| < $ 4.5 \\
		2 jets, $|\eta| < $ 4.5  & 1 jet, $|\eta| < $ 4.5  & 2 jets, $|\eta| < $ 4.5   \\
		1 $b$jet, $|\eta| < $ 2.5 & --- & 1 $b$-jet, $|\eta| < $ 2.5  \\
		%    = 1 \Pqb-jet (77\%) with $|\eta| < $ 2.5 & --- & = 1 \Pqb-jet (77\%) with $|\eta| < $ 2.5 & = 1 \Pqb-jet (77\%) with $|\eta| < $ 2.5 \\
		--- & VR/CR: $m_T(\ell_{W},\nu) > 20/\SI{60}{\GeV}$ & ---  \\
		\bottomrule
	\end{tabular}
		\label{tab:cuts}
		\caption{%
			Overview of the selection applied in the signal, validation and diboson control region.
		}
\end{table}

There are two different types of background sources for this analysis: processes resulting in final states with three real leptons, such as diboson production (mostly consisting of $WZ$ events), associated top-quark pair production with additional bosons ($ttV+ttH$) and $tWZ$, or events in which one of the leptons is non-prompt or is a jet misidentified as a lepton. 
For the first category (with three real leptons), the background contributions are estimated from MC simulations and normalised by their predicted SM cross-sections. Additionally, the diboson normalisation is corrected by a scale factor derived in a dedicated control region (Diboson CR) that is selected as presented in Table \ref{tab:cuts}.

The main sources of events with a non-prompt lepton are the top-quark pair production ($t\bar{t}$) and $Z$ boson production with additional jets. For estimating the correct number of $t\bar{t}$ events, the MC simulation is corrected using a data-MC scale factor derived in a dedicated control region. The $t\bar{t}$ CR is defined using the same cuts as the SR but requiring only one opposite sign, opposite flavour lepton pair and not allowing any OSSF lepton pair, ensuring that there is no contamination from $Z$+jets events. The derived scale factor is $1.21 \pm 0.51$. 

For estimating the number of $Z$+jets events, the fake factor method is used. This is a data-driven technique that relies on calculating fake factors in a region enriched in events containing non-prompt leptons. This region has the same definition as the signal region but $m_T(W)$ is required to be \SI{20}{GeV}. The fake factors are defined as the ratio between events that have all three reconstructed leptons passing tight isolation criteria and events with two tight leptons and one passing looser isolation cuts. These factors, that are derived with respect to the $p_T$ of the lepton associated to the $W$ boson, are then applied to events that pass the SR selection but that have one of the leptons passing looser identification criteria. This is done separately for non-prompt electrons and muons and results in a total $Z$+jets contribution of 37 events, with a $40\%$ assigned uncertainty.

The modelling of all backgrounds is checked by looking at kinematic variables in validation regions enriched in diboson and $t\bar{t}$ events that are defined using the cuts included in \ref{tab:cuts}. Good agreement between the data and MC is observed in all distributions.

\subsection{Neural network training and signal extraction}

In order to separate between signal and background, an artificial neural network (NN) is used to combine the separation power of several observables into a single discriminant. The NN is trained in the SR to separate signal against all backgrounds except for $t\bar{t}$ production (that is discarded due to low statistics in the MC sample, after applying the SR selection). The first variables selected by the network are the $|\eta|$ and $p_T$ of the untagged jet. 

\begin{figure}[htb]
	\centering
	\includegraphics[width=.65\linewidth]{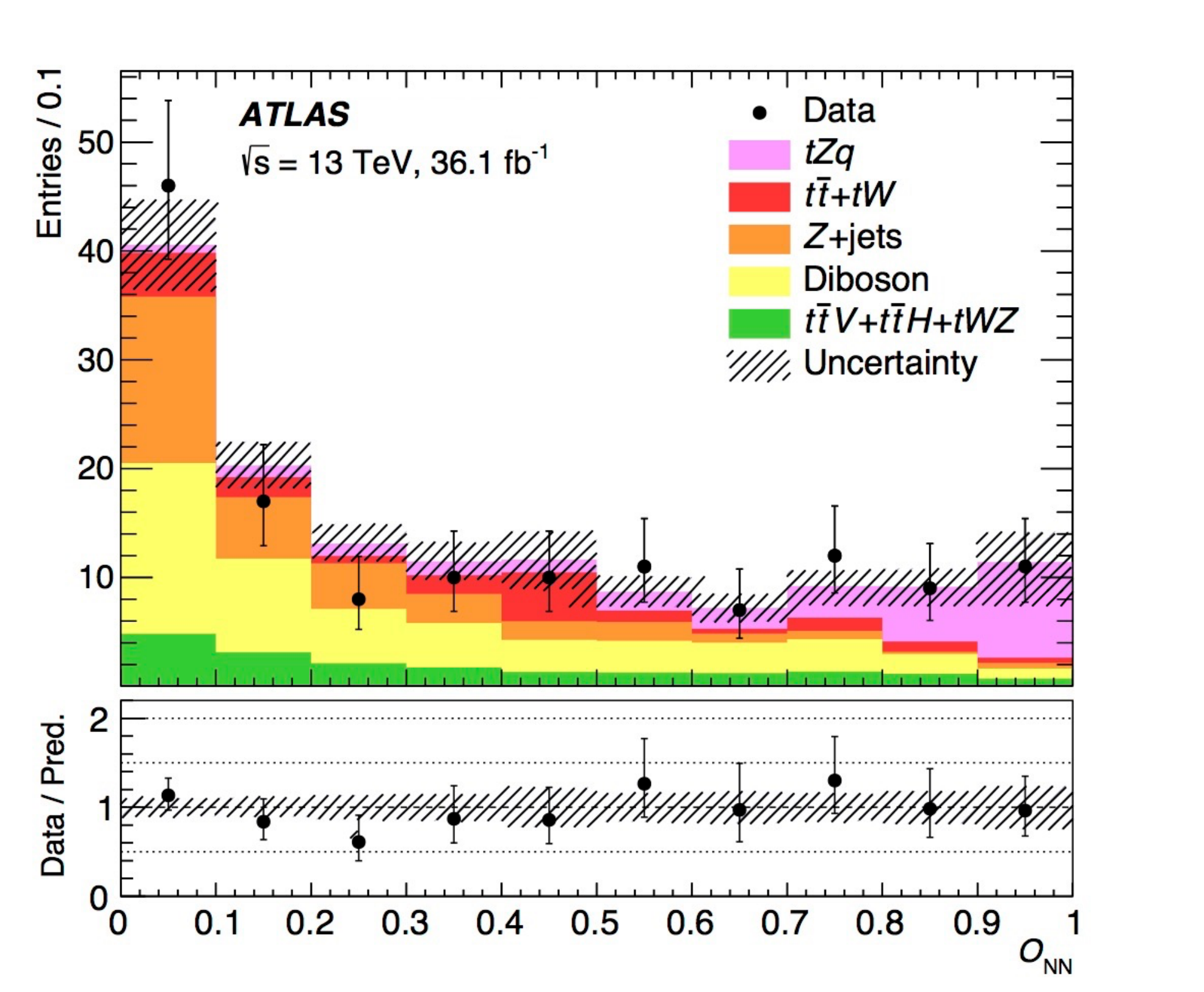}
	\caption{Post-fit NN output distribution in the SR. Signal and backgrounds are normalised to the expected number of events after the fit. The uncertainty band includes both the statistical and systematic uncertainties as obtained by the fit \cite{Aaboud:2017ylb}.}
	\label{fig:NNout}
\end{figure}

The $tZq$ cross-section is extracted by performing a binned likelihood fit of the NN output distribution in the SR. Several sources of systematic uncertainties are taken into account and included as additional nuissance parameters in the fit. The dominant systematic uncertainty is related to the amount of additional radiation in the signal sample. This is evaluated by considering additional samples in which the hard-scatter scales and the scales in the parton shower are simultaneously varied. The result is also affected by large statistical uncertainty (amounting to $28\%$), while the total systematic error is $23\%$.  

In total, $141$ events are observed after applying the SR selection. The total pre-fit number of predicted events for all backgrounds and signal is $162$, with $35$ coming from $tZq$ production. The post-fit number of signal events is $26$.  
Figure \ref{fig:NNout} shows the NN discriminant after normalising the signal and background to the fitted number of expected events. Very good agreement between data and MC is observed. 

%%%%%%%%%%%%%%%%%%%%%%%%%%%%%%%%%%%%%%%%%%%%%%%%%%%%%%%%%%%%%%%%%%%%%%%%%
%%
%%   use this format to include a LaTeX table  into your paper
%%
%\begin{table}[t]
%\begin{center}
%\begin{tabular}{l|ccc}  
%Patient &  Initial level($\mu$g/cc) &  w. Magnet &  
%w. Magnet and Sound \\ \hline
% Guglielmo B.  &   0.12     &     0.10      &     0.001  \\
% Ferrando di N. &  0.15     &     0.11      &  $< 0.0005$ \\ \hline
%\end{tabular}
%\caption{Blood cyanide levels for the two patients.}
%\label{tab:blood}
%\end{center}
%\end{table}
%%%%%%%%%%%%%%%%%%%%%%%%%%%%%%%%%%%%%%%%%%%%%%%%%%%%%%%%%%%%%%%%%%%%%%%%%%%

\section{Results}

The analysis investigating the production of a single top-quark in association with a $Z$ boson was performed using \SI{36.1}{\per\femto\barn} of proton-proton collision data at a centre-of-mass energy of \SI{13}{\TeV}. The measured cross-section: $\sigma_{tZq} = 600\pm 170\text{(stat.)} \pm 140 \text{(syst.) fb}$ is in good agreement with the NLO SM prediction. Evidence for the signal is found with a 4.2 (5.4) $\sigma$ observed (expected) significance. 

\printbibliography[heading=bibintoc]
 
\end{document}